\PassOptionsToPackage{dvipsnames}{xcolor}
\documentclass[conference]{IEEEtran}
\IEEEoverridecommandlockouts
\usepackage{cite}
\usepackage{amsmath,amssymb,amsfonts}
\usepackage{algorithmic}
\usepackage{graphicx}
\usepackage{textcomp}
\usepackage{xcolor}
\usepackage{balance}
\usepackage[all]{nowidow}

\usepackage[hidelinks]{hyperref}
\usepackage[capitalise]{cleveref}

\usepackage[acronym]{glossaries}
\newacronym{osm}{OSM}{OpenStreetMap}
\newacronym{crs}{CRS}{Coordinate Reference System}
\newacronym{pip}{PiP}{Point-in-Polygon}

\usepackage{todonotes}
\definecolor{darkblue}{rgb}{0.0, 0.0, 0.55}
\begin{document}

\title{Spatial Analysis on Value-Based Quadtrees of Rasterized Vector Data}

\author{\IEEEauthorblockN{Diana Baumann, Nils Japke, Tim C. Rese, David Bermbach}
    \IEEEauthorblockA{\textit{Technische Universit\"at Berlin}\\
        \textit{Scalable Software Systems Research Group} \\
        \{diba,nj,tr,db\}@3s.tu-berlin.de}
\thanks{© 2026 IEEE.  Personal use of this material is permitted.  Permission from IEEE must be obtained for all other uses, in any current or future media, including reprinting/republishing this material for advertising or promotional purposes, creating new collective works, for resale or redistribution to servers or lists, or reuse of any copyrighted component of this work in other works.}}


\maketitle

\IEEEpubidadjcol

\begin{abstract}
    Mobility data science offers insights into the complex interconnections of spatial data of moving objects and their surroundings, often based on a combination of vector and raster data. For example, mobility traces are usually in vector format, weather data are often in raster format. Yet, available spatial analysis tools for exploratory data science push data scientists towards one or the other, providing only limited support for the respective other.
    
    In this paper, we contribute to this problem space with a value-based quadtree index, which serves as a bridge builder to support joint spatial analysis on vector and raster data leveraging their unique autocorrelation property. We achieve a 90\% reduction in median \gls{pip} query latency, while keeping the accuracy of query responses at equal level.
\end{abstract}


\section{Introduction}
\label{sec:introduction}
Spatial data are available in vast volumes from a variety of sources and the volume keeps growing continuously.
    Satellite images that cover natural phenomena, green city areas in urban planning, or cycling trajectories in mobility science are just a few examples of the wide range of application domains that work with spatial data.
    Analysis on such data and the combination of different data sources and formats, in particular, provides insights into their complex interconnections~\cite{alam_survey_2022,doulkeridis2021survey}.

Due to the sheer amount and variety of spatial data, challenges such as expensive processing and lack of interoperability of different data formats arise.
    For example, it is a common use case to correlate cycling trajectories with road conditions and weather data.
    While trajectories are often represented as vectors, road condition and weather data may be provided as images or by measurement stations and can either be vector or raster data.
    Correspondingly, the question of how to efficiently process combinations of raster and vector data is an important one.
		
Although popular big data systems exist, they either specialize on spatial vector data~\cite{alarabi2017st-hadoop,xie2016simba,alarabi2020summit} or raster data~\cite{baumann1998rasdaman,scidb} but not both.
    At the same time, stand-alone database systems such as PostGIS\footnote{\url{https://postgis.net}} and MobilityDB\footnote{\url{https://mobilitydb.com}} support spatial data, however, given their different focus, they lack support for data transformation, reprojection, on-demand ingestion, raster operations, and combinations of different \glspl{crs} between vector and raster data~\cite{doulkeridis2021survey}.
    Other approaches tackle interoperability of vector and raster data by either building complex systems that completely omit data transformation, i.e., no rasterization or vectorization~\cite{eldawy2017vectorraster,singla2021raptor}, or work on raster data exclusively~\cite{rdpro2024shang}.
    As such in currently available tools, data scientists are pushed towards either vector or raster data even though exploratory data analysis clearly relies on the intersection of both data formats.

In this paper, we contribute to this problem space with a pipeline that includes a bridge builder to support joint spatial analysis on vector and raster data by leveraging their unique spatial autocorrelation property.
    Due to the unique features regarding spatial dimensions, spatial data often share the fact that neighboring data entries have either the same or a similar value, i.e., they are autocorrelated through spatial attributes~\cite{gisBook}.
    A trajectory of a moving object, e.g., a cycling route, will most likely be represented as a continuous series of data points.
    Similarly, for raster data, a cell inside a park polygon is usually surrounded by neighbors that are also inside the polygon.
    While there are corner cases, the majority of data points naturally share features with their respective neighbors.   
 
\begin{figure}
    \centering
    \includegraphics[width=\columnwidth]{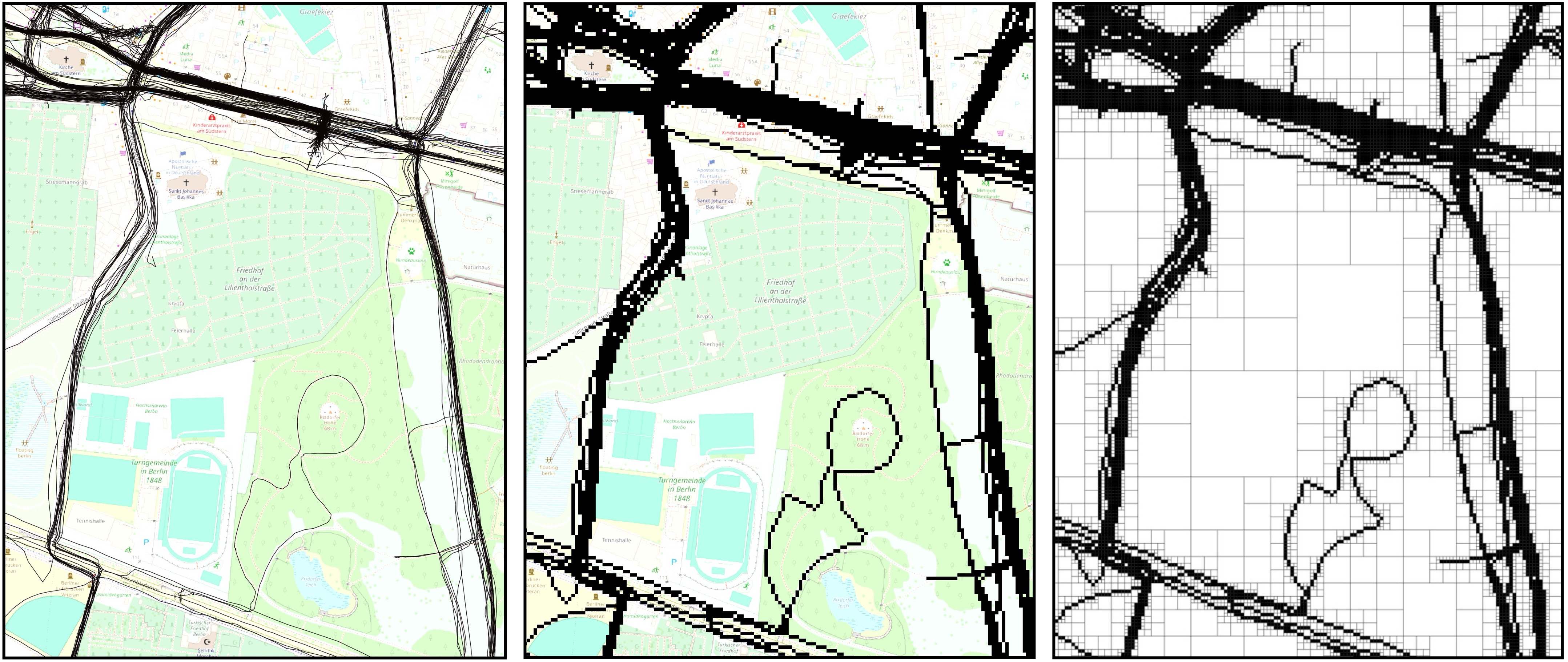}
    \caption{We rasterize the trajectory vector data (left) with a resolution of 5$m^{2}$ to a raster representation (middle) and create a quadtree index (right).}
    \label{fig:v-r-qt}
\end{figure}

Spatial autocorrelation allows us to build a value-based quadtree index~\cite{samet1984quadtree} on public datasets that compresses potentially a huge number of correlated, identical cells and stores them in one single index entry.
    In our approach, it serves as a bridge builder for vector and data intersection supporting both data formats within one exploratory data analysis tool.
    With the organization of the cells into a hierarchical tree structure, an index allows querying the space faster.
    To make the index compatible for non-raster datasets, i.e., vectors, we use one common raster structure as a first step and build a quadtree index on the rasterized data that summarizes pixels with similar values (see~\cref{fig:v-r-qt}).
    With this approach, we address the typical limitation of raster analysis where every raster cell needs to be evaluated regardless of its possible spatial autocorrelation to its neighbors.
    
Compressing the raster into a value-based quadtree allows us to run queries efficiently by skipping non-related cells and minimizing access to autocorrelated cells, which keeps processing demands low and the query accuracy high, all within one joint spatial analysis workflow.
    Furthermore, this approach allows us to analyze all datasets that are needed for a use case by intersecting and bringing them all into the same data structure.

The contributions of our paper are threefold:
\begin{enumerate}
    \item We propose an \texttt{R}-based prototype, which supports combined spatial vector and raster data analysis.
    \item We show that due to the autocorrelation property of spatial data, we compress potentially a huge number of correlated, identical cells and still can run fast and accurate \gls{pip} queries against their quadtree representations, thus, allowing for joint exploratory data analysis (in~\cref{sec:design}).
    \item We evaluate our approach using a prototype implementation and running experiments against it (in~\cref{sec:evaluation}).
\end{enumerate}
\section{Background}
\label{sec:background}
Spatial data are present in various formats, each with their own characteristics and challenges.
    Mobility traces of cyclists, Earth observation data, or satellite movement data are examples of spatial data with different data formats, resolutions, and \gls{crs}~\cite{liang_survey_2024,karakaya2020simra}.
    In this section, we give an overview vector and raster data as well as rasterization and spatial indexing strategies on such data.

\paragraph*{Vector Data}
Spatial vector data can be points, polylines, or polygons:
    Points are geometric objects used to represent individual locations by a pair of coordinates, latitude and longitude.
    Polylines, or line strings, represent continuous lines through a sequence of points, e.g., trajectories of moving objects.
    To represent a curved road, polylines approximate the actual line with a series of points.
    Polygons are used for areas, buildings, or natural landscapes.
    The higher the resolution of vector data, the more resources are required for processing them~\cite{gisBook}.

\paragraph*{Raster Data}
Raster data represent spatial data as a matrix, where the space is gridded and mapped to a corresponding \gls{crs} that pinpoints each cell to a specific area on Earth.
    Trajectories are represented on a raster by flagging each cell through which the trajectory moves.
    While staying consistent over space, rasters do not always support lossless conversion to different resolutions or resampling to a different \gls{crs}.
    Moreover, the latency of queries against raster data increases with higher resolution as all raster cells need to be accessed during the analysis~\cite{teng_efficient_2024}.

\paragraph*{Rasterization}
Vector and raster data can be translated into the respective other format.
    Rasterization, i.e., the conversion of vector objects to raster objects, needs to hold the information about the \gls{crs} and the resolution of the raster~\cite{gisBook}.
    The higher the resolution of a raster, the smaller the area a single cell covers, which leads to improved accuracy during the analysis.
    Rasterization of points is straight-forward: The cell that the point falls into is assigned the value of the point.
    However, when rasterizing a polygon where a border only partially covers a cell, rasterization rules need to define how much coverage is needed in order for the cell to be assigned to the polygon.
    Even with high resolution rasters, this step can inject inaccuracies into the data, which leads to less accurate queries on rasterized vector data.

\paragraph*{PiP Queries and Spatial Indexing Strategies}  
\gls{pip} queries, among other query types, are a common analysis which checks whether a point is contained in one or more polygons of a dataset.
    This, typically, requires checking every candidate polygon in a potentially large dataset.
    Here, spatial indexing can improve efficiency~\cite{indexing2002}; for spatial raster data, this can, e.g., be done through quadtrees~\cite{samet1984quadtree}.
    A quadtree index recursively divides the raster into four quadrants until all cells within a quadrant share the same value.
    Those cells are then summarized by one index entry.
    This way, queries no longer need to check all raster cells, instead only searching in the relevant (sub-)quadrants.
\section{Design}
\label{sec:design}

\begin{figure}
    \centering
    \includegraphics[width=\columnwidth]{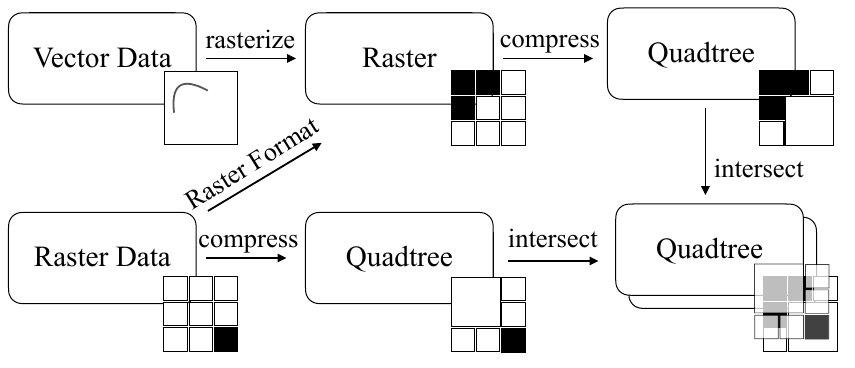}
    \caption{System Overview: The first dataset containing trajectories is rasterized to a 5$m^{2}$ raster, e.g., as the raster dataset to match the same structure. A quadtree index is created on both rasters individually, which can be intersected for faster analysis.}
    \label{fig:design}
\end{figure}

In this section, we show our data analysis pipeline for spatial vector and raster data, which rasterizes the datasets to the same raster structure and creates value-based quadtrees on the datasets needed for analysis (an example is shown in the system's overview in~\cref{fig:design}).
    These datasets can then be intersected and analyzed as one.
    The utilities of all steps mentioned here are provided by the \texttt{R} libraries \texttt{terra}\footnote{\url{https://github.com/rspatial/terra}} except for the quadtree creation which is provided by \texttt{quadtree}.\footnote{\url{https://dfriend21.github.io/quadtree/}}

\paragraph*{Rasterization}
As we stated previously, spatial data come in different formats and \gls{crs} from a variety of sources, which requires some sort of preprocessing.
    As this highly depends on the use case, we do not consider this to be part of the pipeline but show an example in our evaluation later.
    However, in order for the datasets to deliver accurate results during the analysis, they need to be re-projected to a common \gls{crs}. 

We rasterize all datasets so that they follow the same raster structure.
    This can either be set up from scratch or provided from another raster dataset.
    Further, the resolution and \gls{crs} are defined, so that the cells of all rasters cover the same area and area size.

\paragraph*{Value-Based Quadtree Index Creation}
After all datasets are rasterized, we create a quadtree index on the rasters based on the values of the datasets.
    The spatial autocorrelation allows us to compress all respective cells and store them in one single index entry.
    This allows to describe large, spatially autocorrelated areas in fewer cells than in a raster without losing information.
    
\paragraph*{Intersecting the Quadtrees}
Intersecting the datasets, allows us to run accurate queries directly on the intersection of all the datasets that are needed for the analysis. 
    While previous solutions process the datasets separately, we intersect vector and raster data, which results in intersected quadtrees and so support accurate analysis on the intersection.
    We show how this works for two datasets in our evaluation.
    However, our concept supports intersecting any relevant datasets as long as the steps of our approach are followed so that all have the same quadtree structure.

\paragraph*{PiP-Query Analysis}
We extract point values from the quadtree index by accessing the respective entry.
    Compressed rasters as quadtrees allows us to run \gls{pip} queries efficiently by skipping non-related cells and minimizing access to autocorrelated cells.
    With this approach, we avoid the typical limitation of raster analysis where every raster cell needs to be evaluated regardless of its possible spatial correlation to its neighbors.
\section{Evaluation}
\label{sec:evaluation}
In this section, we show the results of our evaluation, which we based on a use case using publicly available datasets.\footnote{Code and data are available in \url{https://github.com/dianabaumann/value-based_quadtrees.git}.}
    We, first, show how our approach performs on single datasets that differ in their spatial autocorrelation using a use case to illustrate its applicability.
    Then, we intersect both datasets and evaluate our prototype on the intersection of the datasets which is what is usually desired when analyzing complex use cases.

\paragraph*{Use Case}
Mobility data science research shows that spatial data can reveal insights far beyond the original purpose. 
    Prior research shows that cycling trajectories, originally collected to detect near-miss incidents, can be used to analyze surface quality of bike lanes~\cite{karakaya2023crowdsensing}, a use case that was not part of the original research purpose.
    The combination with road quality data led to insights that further can be used to learn about the cycling experience of cyclists.
    City planning relies heavily on such insights in order to be able to design cities according to the citizens' best interest, which has already motivated further research towards, e.g., realistic modeling of cyclists behavior.
    As such, we consider cycling trajectories and green areas represented as polygons and build our evaluation around those two datasets.

\paragraph*{Datasets}
To analyze our use case we use two publicly available datasets.
    The first dataset\footnote{The SimRa dataset is publicly available in \url{https://github.com/simra-project/dataset.git}.} is provided by SimRa~\cite{karakaya2020simra}.
    Its original purpose is to track near-miss accidents and, thus, contributing to more safety for cyclists on the roads.
    From this dataset, we include the cycling trajectories only in Berlin from 01/2024 to 07/2024.
    The data is first cleaned and reduced to linestrings of 4,000 cycling trajectories (221,130 points) to one single GeoPackage file.
    
Berlin parks are taken from \gls{osm}\footnote{\url{https:
//www.openstreetmap.org}} and reduced to a set of (multi-)polygons also in a single GeoPackage file.
    Both datasets are reprojected to the standard \gls{crs} 25833 which covers this region completely.
\paragraph*{Experiment Setup}
We evaluate our approach by running a random sample of 100,000 points within the dataset area for \gls{pip} queries against the quadtree, raster, and vector representations of the datasets 30 times measuring the query latency and query accuracy of each run on an Intel Xeon 4310 CPU with 64GB of RAM, running Debian 13.

\paragraph*{Query Latency Evaluation}
\begin{figure}
    \centering
    \includegraphics{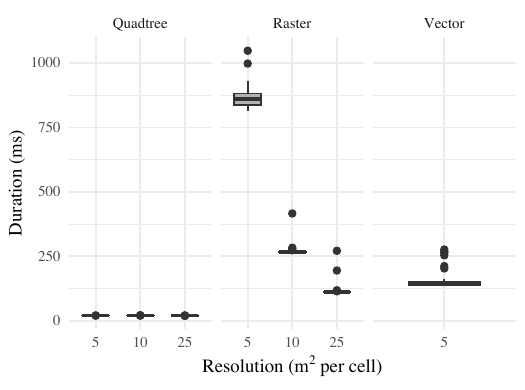}
    \caption{\gls{pip} query duration on three data formats vector, raster, and quadtree sorted by resolution of data with high spatial autocorrelation (parks in Berlin from \gls{osm}). As the resolution can be modified for raster and quadtree data only, we compare different resolutions to the original vector data, indicated as ``5''.}
    \label{fig:autocorrelated}
\end{figure}

We compare the query latency of the \gls{pip} queries for vector, raster and quadtree representations of the Berlin parks dataset.
    We rasterize the vector data in three resolutions of 5, 10, and 25$m^{2}$, and create a quadtree representation on each.
    As depicted in~\cref{fig:autocorrelated}, the query latency on the quadtree is lowest for all resolutions, overall being 10$\times$ faster than the original vector data.
    Raster representation performs worst at high resolution due to the fact that every raster cell needs to be evaluated.
\begin{figure}
    \centering
    \includegraphics{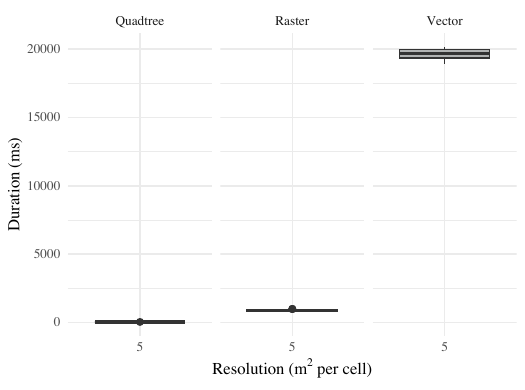}
    \caption{\gls{pip} query duration on the three data formats vector, raster, and quadtree sorted by resolution of data with low spatial autocorrelation (cycling trajectories in Berlin from SimRa).}
    \label{fig:non-autocorrelated}
\end{figure}

We repeat the experiment on a less spatially autocorrelated dataset, the cycling trajectories by SimRa, rasterized to a resolution of 5$m^{2}$.
    In~\cref{fig:non-autocorrelated}, we see that the quadtree still performs best, even though the latency of queries on raster and vector data increases.
    The query latency on the vector format is higher due to the fact that the SimRa dataset consists of line strings, each having up to 1,000 points that need to be evaluated.
    Even with this less spatially autocorrelated dataset, the quadtree representation keeps performing by far better than vector.
    The difference of latency between the raster in the first experiment and here occurs due to the smaller dataset size, i.e., fewer cells that need to be evaluated in this second dataset.
\begin{figure}
    \centering
    \includegraphics{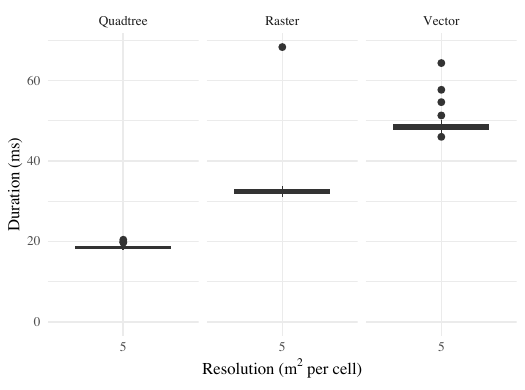}
    \caption{\gls{pip} query duration on the three data formats vector, raster, and quadtree on the intersection of both datasets, SimRa and Berlin parks.}
    \label{fig:latency-intersection}
\end{figure}

\paragraph*{Intersecting Vector and Raster Data}
In particular, our approach allows us to analyze our use case more efficiently.
    As our use case requires two datasets, parks and cycling trajectories, we can bring both datasets into the same raster structure, intersect, and create the quadtree representation on the intersection.
    This allows us to run \gls{pip} queries against both datasets at the same time instead separately, e.g., does the point intersect with a cycling trajectory \emph{and} a park?
    We repeat our query latency experiments on the intersection of the datasets and summarize the results in~\cref{fig:latency-intersection}.

The quadtree representation performs best (approx. by 40\% compared to vector data), followed by the raster and vector representations.
    The overall latency is lower than during the previous experiments due to the intersection of the datasets being smaller than the original datasets before intersection.

\paragraph*{Query Accuracy Evaluation}
We measure the accuracy after each experiment run by comparing the results of the query on raster and quadtree representations with our baseline, the original vector data.
    Every result that is identical to the one of the vector representation is considered accurate as true positive (TP) or true negative (TN).
    The results are summarized as the ratio of the sum of all correctly identified queries (TP+TN) to all queries issued in the experiment (TP+TN+false positives (FP)+false negatives (FN)) in~\cref{tab:accuracy}.

As the resolution decreases, the accuracy of the \gls{pip} query results also decreases.
    This is especially relevant for the points that fall directly on the edge of a polygon (border of a park).
    With lower resolution, raster and quadtrees may define a cell as completely inside a polygon even though the actual park only partially overlaps with the cell.
    This has to be defined at the rasterization step, though, usually an overlap of more than 50\% of the cell with a park would be considered as fully inside the park, which what we also do.
    To verify this, we repeat the experiment with a random sample of 100,000 \gls{pip} queries drawn three times, with the difference that the points fall on the park borders only.
    For lower resolutions, the accuracy scores drop rapidly as expected and quadtrees equally to rasters.

\begin{table}
    \caption{Accuracy scores for vector data in comparison to its raster and quadtree representations in four resolutions.}
    \centering
    \begin{tabular}{| c || c | c| c| c|} \hline
      Resolution   & \multicolumn{2}{c|}{Full Detail} & \multicolumn{2}{c|}{Borders only} \\
     & Raster & Quadtree & Raster & Quadtree \\ \hline \hline
    5~$m^{2}$ & 100\% & 100\% & 100\% & 100\% \\ \hline
    25~$m^{2}$ & ~97.7\% & ~97.7\% & ~49.4\% & ~49.4\% \\  \hline
    100~$m^{2}$ & ~92.9\% & ~92.9\% & ~0\% & ~0\%  \\ \hline
    \end{tabular}
    \label{tab:accuracy}
\end{table}
\section{Discussion}
\label{sec:discussion}
Our approach allows joint spatial analysis on vector and raster data providing support of both data formats in the exploratory data science: We rasterize and compress high-resolution vector data to value-based quadtree index by using the autocorrelation property of spatial data.
    Though indices are a well-known tool to speed up data lookups in general, our approach takes one step further and shows that this common structure among used datasets can strategically be applied to spatial data analysis due to its unique autocorrelation features.
    Our evaluation shows that the query latency on the quadtree representation decreases while the accuracy of query responses stays equally high even for critical \gls{pip} queries that target borders of polygons.
    Bringing two datasets into the same intersected quadtree structure allows us to query both datasets efficiently.
    Even though we did not encounter any limitations not mentioned here, an overhead evaluation of the preprocessing and indexing steps would support our approach with further insights into its applicability.
    
Considering our chosen use case, data from more sources can be added to the evaluation, e.g., rain data or road conditions.
    Adding more datasets into the pipeline can evaluate the practicability of our approach more realistically, as complex use cases require many different datasets.
    Both datasets used in the evaluation are vector data originally.
    Even though this does not compromise the promise of vector+raster data analysis, as our approach rasterizes both to a common raster structure, other datasets, available as rasters originally would support the applicability of our approach even more.

Our tech stack and library choices, in particular, turn out to set performance limitations to our approach not further evaluated in this work.
    The \texttt{quadtree} library imposes internal limits on the number of data points it can process when creating the index. 
    Large datasets result in out-of-memory errors because all data must be loaded and processed within a single process, and the library does not support distributed or parallel execution.
    Thus, the scalability issue of the library and lack of alternatives sets limitations on how scalable our approach is overall.

Similar to the spatial autocorrelation, temporal features can also be used for quadtrees on data cubes, i.e. spatial rasters over time.
    Moving object data typically includes temporal features, which we plan to enhance our approach to support spatiotemporal data as another open research direction.

Further, while relevant, \gls{pip} queries are still only a subset of spatial queries that are popular in spatial analysis.
    As such, a quadtree library that does not support further query types such as range, kNN, and similarity queries cannot provide insights into how good our approach supports those other query types and, thus, a wider range of use cases.
    For high resolution datasets, we aim to leverage the spatial autocorrelation property to distribute and parallelize the quadtree index creation.

Lastly, even though \texttt{R} provides a multitude of supporting libraries for spatial data analysis, a comparison to similar tools in other programming languages such as Python can provide even further insights into the applicability of our approach and possibly address some of the limitations we mentioned earlier.
\section{Related Work}
\label{sec:related-work}
Research on interoperability of vector and raster data has many facets.
    Systems, such as~\cite{alarabi2017st-hadoop,xie2016simba,alarabi2020summit} specialize on spatial data formats and~\cite{baumann1998rasdaman,scidb} on raster data, but no approaches support both or the combination of both, in particular.
    Stand-alone database systems such as PostGIS and MobilityDB support large amounts of spatial data supporting either vector or raster data but not the combination of both.
    Features such as support of data transformations, reprojection, on-demand data ingestion, raster operations, or combination of different \gls{crs}~\cite{doulkeridis2021survey} on said combination of vector and raster data are a rare find.
    
Other approaches tackle interoperability of vector and raster data, by building complex systems that completely omit data transformation, i.e., no rasterization or vectorization.
    The authors in~\cite{singla2021raptor}, similar to the sort-merge join in relational databases, design their approach to handle raster and vector data without the pre-processing step, i.e., vectorization or rasterization.
    Instead of processing raster as is, which would require high processing power with higher resolution, they work with the metadata of the raster and create a hybrid data file containing vector/raster intersections.
    They, however, highlight that the size of the vector dataset remains the limiting factor.
    
\cite{eldawy2017vectorraster} also avoids transformations attempting to process vector and raster data with the \emph{scanline} method.
    It is inspired by algorithms from the graphics community, identifies relevant pixels from the raster data and ignores the others.
    While this initial idea seems promising for handling big spatial data, further evaluation remains subject to future work.
    
Further related research towards efficient processing of vector and raster data is conducted by~\cite{teng_efficient_2024}.
    The authors design an approach which allows querying hybrid (vector and raster) databases by rasterization.
    The raster resolution depends on the features of the geometries, e.g., the number of edges of the polygon.
    This results in one raster per polygon with multiple rasters potentially having different resolutions and alignments.
    In our approach, however, we do not consider the specifics of the features that are being rasterized.
    Instead, we support arbitrary resolution of our raster while keeping the spatial relation of the features.

Some significant work focuses on scalable handling of spatial raster data by~\cite{rdpro2024shang}.
    The authors propose to divide rasters into \emph{Maplets}.
    Unlike tiles, maplets represent one raster but are not tied to it.
    This allows to process big rasters in a distributed manner.
    While innovative and scalable, this approach does not address how maplets from spatial rasters can be interconnected with vector data while remaining the contextual relation.
    
Research on indices for spatial data shows promising ideas towards scalable query processing on raster data as well.
    Zhang et al. chose a quadtree-based data structure for indexing spatial raster data~\cite{zhang2015quadtree}. 
    In their following work, they use bitmap encoding of geospatial rasters into a quadtree index.
    The evaluation clearly shows storage benefits and support for fast (de)compression of large rasters.
    However, the approach is not connected to data analysis of spatial raster data.
\section{Conclusion}
\label{sec:conclusion}
Mobility data science offers insights into the complex interconnections of spatial data of moving objects and its surroundings, and, more often than not, a combination of different data sources and formats is needed.
    Challenges, such as expensive processing and lack of interoperability, arise due to the sheer amount and variety of spatial data.
    Analysis on vector and raster data, in particular, is both common and challenging.
    Yet, available spatial analysis tools for exploratory data science push data scientists towards one or the other, providing only limited support for the respective other.

In this paper, we proposed a bridge builder based on \texttt{R}, which supports vector and raster data intersection.
    We rasterized vector data and leveraged the spatial autocorrelation to compress the data in a value-based quadtree index.
    We, further, intersected related datasets into the same quadtree structure and analyzed them simultaneously.
    Our prototype implementation was able to achieve a 90\% reduction in median \gls{pip} query latency while keeping the accuracy of query response at equal level on high resolution data.
    In the future, we are planning to expand the prototype with support of spatiotemporal data, other query types, intersection of multiple datasets for more complex analysis.

\balance

\bibliographystyle{IEEEtran}
\bibliography{bibliography}

\end{document}